\documentclass[paper,onecolumn]{revtex4-2}
\usepackage[T1]{fontenc}
\usepackage{graphicx,xcolor}
\usepackage{slashed}
\usepackage{float}
\usepackage{verbatim}
\usepackage{subcaption}
\usepackage{bm}
\usepackage{amssymb,amsmath,amsfonts, mathrsfs}

\usepackage{tikz}
\usepackage{tikz-feynman}
\usepackage{tikz-feynhand}

\begin{document}
\title{On the pole trajectory of the subthreshold negative parity nucleon with varying pion masses}


\author{Qu-Zhi Li}
\email{liquzhi@scu.edu.cn}
\author{Zhiguang Xiao}
\email{xiaozg@scu.edu.cn}
\author{Han-Qing Zheng}
\email{zhenghq@pku.edu.cn}
\affiliation{Institute for Particle and Nuclear Physics, College of Physics, Sichuan University, Chengdu  610065, P.~R.~China}



%

\begin{abstract}
	We study the pole trajectory of the recently established subthreshold negative parity nucleon pole, namely the $N^*(920)$, with varying pion masses, in the scheme of linear $\sigma$ model with nucleons using the $N/D$ unitarization method. We find that as the pion mass increases, the pole moves toward the real axis.  For larger pion masses, at tree level, the pole falls to a specific point on $u$-channel cut and crosses to the adjacent Riemann sheet defined by the  logarithmic  $u$ channel cut. At one-loop level, the pole does not meet the $u$-cut up to $m_\pi=0.36$GeV. We also re-examined the $\sigma$ pole trajectory and find it in good agreement with Roy equation analysis result.
\end{abstract}

\maketitle
\section{Introduction}
A subthreshold pole with quantum number $J^{PC}=1/2^-$, named $N^*(920)$, has been
established in the $S_{11}$ channel of $\pi N$  scatterings. It
has been  firstly suggested  in analyzing $\pi N$ scattering data~\cite{ Wang:2017agd,Wang:2018nwi} using the product representation
for partial wave amplitudes  (PWAs)~\cite{Zheng:2003rw,Zhou:2006wm,Zhou:2004ms}, which comes from the correct treatment of the left hand cuts and unitarization~\cite{Xiao:2000kx,He:2002ut}. The pole is also confirmed  using naive
$K$-matrix approach~\cite{Ma:2020sym} and  $N/D$ method~\cite{LiQuZhi:2021nrq}.
Its existence is firmly established in a Roy-Steiner equation analysis 
in Ref.~\cite{Cao:2022zhn}, with the pole location  $\sqrt{s}= (918 \pm 3)-i(163 \pm 9)\mathrm{MeV}$, and is reconfirmed  
in Ref.~\cite{Hoferichter:2023mgy} at $(913.9\pm1.6) -i(168.9\pm3.1)\mathrm{MeV}$. Its properties in turn naturally  become a subject of research interest as  much is left to be  desired.

The early work on the properties of $N^*(920)$ is its coupling to
$N\gamma$ and  $N \pi$~\cite{Ma:2020hpe}. It is found that its
coupling to $N \pi$ is considerably larger than that of
$N^*(1535)$, while its coupling to $N\gamma$ is comparable to that
of $N^*(1535)$.  Here in this paper we will focus on the $N^*(920)$ pole trajectory with varying pion masses.

In the literature, the $\sigma$ pole trajectory with varying $\pi$ masses has been a rather hot topic for discussions,
see for example Refs.~\cite{Hanhart:2008mx,Gao:2022dln} and references therein. Remarkably a model independent Roy equation analysis has been carried out to thoroughly solve the issue~\cite{Cao:2023ntr}. The study of the $\sigma$ pole trajectory with varying $m_\pi$ is important, since it opens a new window in exploring  non-perturbative strong interaction physics provided by lattice QCD calculations. An alternative  study based on $O(N)$ linear $\sigma$ model~\cite{Lyu:2024lzr,Lyu:2024elz} finds similar results comparing with that of Ref.~\cite{Cao:2023ntr}, hence providing further evidence that the $\sigma$ meson may be more reasonably described as “elementary”, in the sense that it is, the same as pions,  described by an explicit field degree of freedom in the effective chiral lagrangian\footnote{Early studies using large $N_c$ (number of colors) arguments also support such a
	suggestion~\cite{Guo:2006br,Guo:2007ff,Guo:2007hm}. }. Inspired by this, we in this paper adopt the
effective lagrangian with a linearly realized chiral symmetry. To be specific we use the renormalizable   toy linear $\sigma$ model with nucleon fields, though it is known that renormalizability condition is not at all a physical requirement when describing low energy hadron physics. In short, we use linear $\sigma$ model rather than the $\chi$PT lagrangian, mainly for  theoretical considerations\footnote{For example, at high temperatures, it is easy to restore the $O(4)$ symmetry using linear $\sigma$ model. On the contrary, in the $\chi$PT framework,   such a restoration is still absent~\cite{Lyu:2024elz, Lyu:2024lzr}.  }, though we believe the two give more or less the same results at qualitative level. 

In the following we begin by a brief introduction of the linear $\sigma$ model with nucleons in Sec.~\ref{sigma}, and calculate the $\sigma$ pole trajectory using [1,1] Pad\'e approximation in Sec.~\ref{N920}. As it is verified that the unitarity approximation does give a similar $\sigma$
pole trajectory as comparing with that of Ref.~\cite{Cao:2023ntr}, it is satisfactory to use the same approximation method to further explore the $N^*(920)$ trajectory, which will also be discussed in Sec.~\ref{N920}. Sec.~\ref{disc} is devoted to discussions and conclusions.

\section{A brief review of linear $\sigma$ model}\label{sigma}


The linear $\sigma$ model~\cite{Gell-Mann:1960mvl} (LSM) lagrangian with a nucleon field can be written as follows:
\begin{equation}
	\begin{aligned}\label{LinearS}
		\mathcal{L}  = & \bar{\Psi}_0i \gamma^\mu \partial_\mu\Psi_0-g_0 \bar{\Psi}_0\left(\sigma_0{+i }\gamma_5
		\vec{\tau} \cdot \boldsymbol{\pi}_0\right)
		\Psi                                                                                                                                                         \\
		               & + \frac{1}{2}\left(\partial_\mu \sigma_0 \partial^\mu \sigma_0+\partial_\mu \boldsymbol{\pi}_0 \cdot \partial^\mu \boldsymbol{\pi}_0\right)
		-\frac{\mu_0^2}{2}\left(\sigma_0^2+\boldsymbol{\pi}_0^2\right)-\frac{\lambda_0}{4 !}\left(\sigma_0^2+\boldsymbol{\pi}_0^2\right)^2
		+C \sigma_0\ ,
	\end{aligned}
\end{equation}
where $\Psi_0$ is the isospin doublet denoting bare nucleon fields, and
$\boldsymbol{\pi}_0,\sigma_0,\mu_0, g_0, \lambda_0$ are bare $\pi$ meson
triplet, $\sigma$
field, a mass parameter, and couplings, respectively.  The renormalized
quantities are related to bare ones through:
\begin{equation}
	\left\{\begin{array}{l}
		\psi_0=\sqrt{Z_\psi} \psi\ ,                                                         \\
		\left(\sigma_0, \boldsymbol{\pi}_0\right)=\sqrt{Z_\phi}(\sigma, \boldsymbol{\pi})\ , \\
		\mu_0^2=\frac{1}{Z_\phi}\left(\mu^2+\delta \mu^2\right)\ ,                           \\
		g_0=\frac{Z_g}{Z_\psi \sqrt{Z_\phi} }g\ ,                                            \\
		\lambda_0=\frac{Z_\lambda}{Z_\phi^2} \lambda\ .
	\end{array}\right.
\end{equation}

Spontaneous chiral symmetry breaking ($\chi$SB) occurs when the
$\sigma$ vacuum  expectation value (vev)
$\langle \sigma \rangle =v \neq 0$, generating  three zero-mass Goldstone bosons: $\pi^i,\quad i=1,2,3$
in the absence of explicit $\chi$SB term $C\sigma_0$. To perform a perturbative calculation,
one shifts $\sigma\to\sigma + v$ such that  $\langle \sigma \rangle  = 0 $
and gets,
\begin{equation}\label{Lcoun}
	\begin{aligned}
		\mathcal{L} & =\bar{\psi}\left[i \slashed \partial-m_N -g \left(\sigma+i \pi \cdot \tau \gamma_5\right)\right] \psi \\
		            & +\bar{\psi}\left[-\delta m_N-\delta g
			\left(\sigma+i \pi \cdot \tau
			\gamma_5\right)+i\left(Z_\psi-1\right) \slashed
		\partial\right] \psi                                                                                                \\
		            & +\frac{1}{2}\left[\left(\partial_\mu
			\sigma\right)^2+\left(\partial_\mu \pi\right)^2-m_\sigma^2
		\sigma^2-m_\pi^2 \pi^2\right.                                                                                       \\
		            & \left.+(Z_\phi-1)\left(\left(\partial_\mu
			\sigma\right)^2+\left(\partial_\mu
			\pi\right)^2\right)-\delta m_\pi^2\pi^2-\delta m_\sigma^2\sigma^2\right]-\frac{\lambda}{4!}  \left[\sigma^4 +
		\pi^4 +4v\sigma(\sigma^2+\pi^2) +2\sigma^2\pi^2  \right]                                                            \\
		            & - \frac{\lambda}{4!} (Z_\lambda-1) \left[\sigma^4 +
			\pi^4 +4v\sigma(\sigma^2+\pi^2) +2\sigma^2\pi^2  \right]-\sigma
		\left[ v(m_\pi^2+\delta m_\pi^2) -C
			\sqrt{Z_\phi}
			\right]~ ,
	\end{aligned}
\end{equation}
with
\begin{equation}
	\begin{split}
		 & m_N=gv,\quad m_{\sigma }^2 =   \mu^2 + \frac{1}{2}\lambda v^2 , \quad   m_{\pi }^2 =   \mu^2 + \frac{1}{6} \lambda v^2, \\
	\end{split}
	\label{eq:mass}
\end{equation}
and
	{the renormalization constants are defined as}
\begin{equation}\label{delm}
	\left\{\begin{array}{l}
		\delta m_N= m_N(Z_g-1)\ ,                                            \\
		\delta g=g  (Z_g -1)\ ,                                              \\
		\delta m_\pi^2 = \delta \mu^2+\frac{1}{6}(Z_\lambda-1)\lambda v^2\ , \\
		\delta m_\sigma^2 \equiv \delta \mu^2+\frac{1}{2}
		(Z_\lambda-1)\lambda v^2\ .
	\end{array}\right.
\end{equation}
From Eq.~\eqref{eq:mass} we obtain the relation between $m_\pi$  and $m_\sigma$:
\begin{equation}
	m_\sigma^2 =  m_\pi^2 + \frac{1}{3} \lambda v^2 ~.
	\label{eq:msigma-mpi-v}
\end{equation}
This relation holds if the renormalization constant $Z_\lambda$ is chosen as
\begin{equation}
	Z_\lambda = 1 - \frac{3 (\delta m_\pi^2-\delta
		m_\sigma^2)}{\lambda v^2}~ .
\end{equation}

The other renormalization constants are determined by the following
conditions, as done in Ref.~\cite{Mignaco:1971yr}:
\begin{itemize}
	\item To determine $\delta m_\pi^2$ and $Z_\phi$, we demand  that the full
	      $\pi$ propagator $\Delta_\pi(s)$ satisfies
	      \begin{equation}
		      \begin{split}
			       & i \Delta_\pi^{-1}(m_\pi^2) = 0\,,                                             \\
			       & \left. i \frac{\mathrm{d}\Delta_\pi^{-1}(s)}{\mathrm{d}s} \right|_{s=m_\pi^2}
			      =1      \,.
		      \end{split}\label{eq:Ren-cond-1}
	      \end{equation}
	\item $\delta m_\sigma^2$ can be determined by requiring the real
	      part of the $\sigma$ propagator, $i \Delta^{-1}_{\sigma}(s)$, to vanish when
		      { $s\to m_\sigma^2$,} i.e.,
	      \begin{equation}
		      \operatorname{Re}[i \Delta_\sigma^{-1}(m_\sigma^2)] = 0~.
	      \end{equation}
	      Notice that the parameter $m_\sigma$ can not be identified
	      as the $\sigma$ pole mass when $m_\sigma>2m_\pi$ since $i \Delta_\sigma^{-1}(m_\sigma^2)
	      $ is complex in this situation. 
	      On the other hand, if $m_\pi$
	      increases to be large enough such that $m_\sigma < 2m_\pi$, then  $i
		      \Delta_\sigma^{-1}(m_\sigma^2) $  becomes real and
	      $m_\sigma$ is just the pole mass.
	\item $Z_\psi$ and  $Z_g$ are determined by forcing the full nucleon
	      propagator $\Delta_N(\slashed{p})$ behaving like:
	      \begin{equation}
		      \begin{split}
			       & i \Delta_N^{-1}( m_N) = 0\ , \\
			       & i \frac{\mathrm{d}
				      \Delta_N^{-1}(\slashed{p})}{\mathrm{d}\slashed{p}}
			      \big|_{\slashed{p}=m_N} = 1\ .
		      \end{split}
	      \end{equation}

\end{itemize}
The results for the renormalization constants and counter terms
under these renormalization conditions are listed in  Appendix~\ref{ece}.

In LSM, there are four free parameters and they can be chosen as $\lambda$,  $m_\sigma$, $m_\pi$,
$g$. From Eqs.~\eqref{eq:msigma-mpi-v} and~\eqref{eq:mass}, the vev $v$ and the  nucleon mass  $m_N$ are expressed by:
\begin{equation}\label{lambda}
	v^2= \frac{3(m_\sigma^2-m_\pi^2)}{\lambda}, \quad  m_N =  gv.
\end{equation}
In the physical situation, $m_\pi = 0.138\mathrm{GeV},
	m_N=0.938\mathrm{GeV}$, and $v$ is identical to the pion decay constant
$f_\pi$ at tree level, whose experimental value  is $0.093 \mathrm{GeV} $ so
that $g\simeq 10$. From PCAC, one also obtains
$C\sqrt{Z_\phi}=f_\pi m_\pi^2$, and the pion decay constant $f_\pi$ are
related to $v$ by~\cite{Lee:1968da}:
\begin{equation}\label{vfpi}
	v = f_\pi m_\pi^2 i \Delta_\pi(0)~.
\end{equation}
At the one-loop level, this equation reads:
\begin{equation}
\begin{split}
    	    \frac{\lambda^2v^2 }{144\pi^2}&\left[
		\left(B_0(m_\pi^2,m_\pi^2,m_\sigma^2)-
		B_0(0,m_\pi^2,m_\sigma^2)\right)/m_\pi^2-B^\prime_0(m_\pi^2,m_\pi^2,m_\sigma^2)\right]\\
        &-
	\frac{g^2m_\pi^2}{4
		\pi ^2}B^\prime_0(m_\pi^2,m_N^2,m_N^2)
	=
	\frac{f_\pi }{v}-1~.
\end{split}
\end{equation}
 Numerical tests reveal that the left hand side of this equation has a
 negligible effect within the parameter regime used in our following
 calculations~\cite{Mignaco:1971yr}. Therefore,  the $f_\pi$  is still
 approximately  equal to the vev $v$ at the one-loop level~\footnote{We also
 tested fixing $v=f_\pi$ at the physical value without $m_\pi$ dependence and
 found that this did not modify the pole trajectories too much. Thus, the pole
 trajectories are in fact quite insensitive to the $m_\pi$ dependence of $v$.}. 

Since our purpose of using the LSM is to approximate  QCD which has
only two free parameters, i.e., gauge coupling or $\Lambda_{QCD}$ and the quark mass,
whereas LSM has four --- $(g,m_N,m_\sigma,m_\pi)$, this leaves some room for manipulation of $m_\pi$ dependence of parameters. 
Therefore, we make use of $\chi$PT results to fix the dependence of $f_\pi$  on $m_\pi$\footnote{This may be somewhat equivalent to adding some high order contributions in LSM lagrangian. }, which  has been calculated up to NNLO in Refs.~\cite{Bijnens:1997vq,Bijnens:1998fm}.  Rewriting in terms of $F$ and $m_\pi$, $f_\pi$ is expressed as~\cite{Bijnens:1997vq,Bijnens:1998fm}
\begin{equation}
	\begin{split}
		 & f _ { \pi } = F \Biggl [ 1 + F _ { 4 } \frac { m _ { \pi } ^ { 2 } } { 1 6 \pi ^ { 2 } F ^ { 2 } } + F _ { 6 } \biggl ( \frac { m _ { \pi } ^ { 2 } } { 1 6 \pi ^ { 2 } F ^ { 2 } } \biggr ) ^ { 2 } \Biggr ] , \quad F _ { 4 } = 1 6 \pi ^ { 2 } l _ { 4 } ^ { r } - \log \frac { m _ { \pi } ^ { 2 } } { \mu ^ { 2 } }~,                                                                                                                                                                                   \\
		 & F _ { 6 } = \left( 1 6 \pi ^ { 2 } \right) ^ { 2 } r _ { F } ^ { r } - 1 6 \pi ^ { 2 } \left( l _ { 2 } ^ { r } + \frac { 1 } { 2 } l _ { 1 } ^ { r } + 3 2 \pi ^ { 2 } l _ { 3 } ^ { r } l _ { 4 } ^ { r } \right) - \frac { 1 3 } { 1 9 2 } +\\
         &\quad\quad\left( 1 6 \pi ^ { 2 } ( 7 l _ { 1 } ^ { r } + 4 l _ { 2 } ^ { r } - l _ { 4 } ^ { r } ) + \frac { 2 9 } { 1 2 } \right) \log \frac { m _ { \pi } ^ { 2 } } { \mu ^ { 2 } } - \frac { 3 } { 4 } \log ^ { 2 } \frac { m _ { \pi } ^ { 2 } } { \mu ^ { 2 } } ,
	\end{split}
\end{equation}
where these low energy constants (LECs) are obtained in Ref.~\cite{Niehus:2020gmf} by a full analysis on lattice data. In the following calculations,  we impose this condition to constrain the $m_\pi$  dependence of $v$ by using Eq.~\eqref{vfpi}.  Another constraint from $\chi$PT is the $m_\pi$ dependence of $m_N$. We require the dependence of  $m_N$ on $m_\pi$  to match the $\mathcal{O}(p^5)$ result of  baryon $\chi$PT~\cite{Liang:2025cjd}. From the second equation in \eqref{lambda}, the $m_\pi$ dependence of $g$ is also fixed. With one remaining free parameter, from the first equation in \eqref{lambda}, we consider two alternative simple assumptions to proceed:  to fix $m_\sigma$
such that $\lambda$ is dependent on $m_\pi$ or vice versa. Since these two choices produce similar  qualitative results, we choose
the first one in discussing the $\pi\pi$ scatterings and only provide the results in $\pi N$ scatterings for both choices.

\section{The trajectories of the  $\sigma$ pole and the $N^*(920)$ pole}\label{N920}

This section is devoted to the study of the $\sigma$ pole and the $N^*(920)$ pole dependence
on varying pion masses. This analysis is meaningful in understanding the non-perturbative aspects of low energy
strong interaction physics, especially in the era when lattice QCD studies become more and more prosperous\footnote{For a recent review to the related subjects, one is referred to Ref.~\cite{Rodas:2023nec}.}.   For the former, the $\sigma$ pole trajectory is well understood~\cite{Cao:2023ntr,Rodas:2023gma} while knowledge of $N^*(920)$ trajectory  is absent yet, from either  lattice QCD  or analytical studies.

We will study the two trajectories based on the renormalizable linear
$\sigma$ model with nucleons. The reason why we choose such a model is
already discussed in the introduction. We in the following firstly
re-analyze the $\sigma$ pole trajectory using  [1,1] Pad\'e
approximation\footnote{It is not at all obvious that Pad\'e
	approximation leads to a satisfied solution since it may be spoiled by spurious poles~\cite{Qin:2002hk}.
	But in linear $\sigma$ model it works rather good, since crossing symmetry is preserved at least partially.}. It will be found that the trajectory obtained is in rather
good agreement with that of Roy equation analyses and $O(N)$ model
results qualitatively. This is because, unlike the unitarization using $\chi$PT amplitude, the $t$ and $u$-channel $\sigma$ exchange is  taken into account here, which is the requirement of crossing symmetry.

\subsection{The $\sigma$ pole location in the $I,J=0,0$ channel  $\pi\pi$ scattering amplitude}

Basically there can be two ways to extract the $\sigma$ pole location: one is from the $\sigma$ propagator, another  is from the unitarized $\pi\pi$ scattering amplitude. They are not equivalent under the approximations being used, however. The propagator is obtained by using  a Dyson resummation of self energy bubble chain, and is essentially a one loop calculation, whereas the pole in the unitarized amplitude contains more complete dynamical input\footnote{To be more specific, at one loop, the self energy on the second sheet contains the pseudo-threshold but not the dynamical left hand cut (the crossed channel $\sigma$ exchanges), which is presented anyway in unitarized scattering amplitudes. So the two solutions cannot be the same. }. Therefore we adopt the scattering amplitude to extract the pole locations.

The $\pi\pi$ elastic scattering amplitude is written as:
\begin{equation}
	T(s,t,u) = A(s,t,u)\delta_{\alpha\beta}\delta_{\gamma\delta} +
	B(s,t,u)\delta_{\alpha\gamma}\delta_{\beta\delta}+
	C(s,t,u)\delta_{\alpha\delta}\delta_{\gamma\beta}
\end{equation}
where  $\alpha,\beta,\gamma$ and  $\delta$ are isospin indexes and
$s,t$ and  $u$ are Mandelstam variables subject to the constraints $s+u+t=4m_\pi^2$.
$A(s,t,u),B(s,t,u)$ and  $C(s,t,u)$ are Lorentz invariant amplitudes.
The total  isospin  $I=0$ amplitude $T^0(s,t,u)$ can be derived as
\begin{equation}\label{T0}
	\begin{split}
		 & T^0(s,t,u) = 3A(s,t,u) +B(s,t,u)+C(s,t,u)~.
	\end{split}
\end{equation}

Partial wave amplitude(PWA) is defined as
\begin{equation}\label{pwj}
	T^{I}_J(s) = \frac{1}{32\pi
		(s-4m_\pi^2)}\int^0_{4m_\pi^2-s}P_J\left(1+\frac{2t}{s-4}\right)T^I(s,t,u)~,
\end{equation}
where $P_J$ is the Legendre polynomial. The elastic unitarity reads:
\begin{equation}
	\operatorname{Im} {T^{I}_J(s)} =
	\rho(s,m_\pi,m_\pi)|T^{I}_J(s)|^2,\quad s > 4m_\pi^2~,
\end{equation}
with
\begin{equation}
	\rho(s,m_1,m_2) = \frac{\sqrt{ (s-(m_1+m_2)^2)(s-(m_1-m_2)^2) }}{s}~.
\end{equation}

\begin{figure}[H]
	\centering    \includegraphics[width=0.8\linewidth]{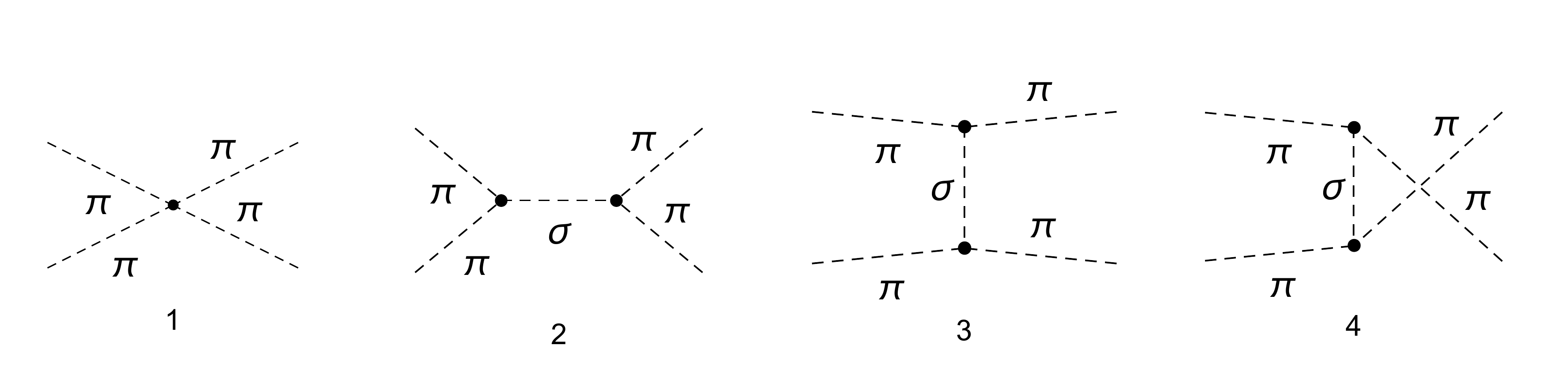}
	\caption{The tree-level Feynman diagrams contributing to $\pi\pi$ scatterings.}
	\label{pipitree}
\end{figure}

The PWA  has been
calculated up to one-loop level within LSM neglecting nucleon
contributions. At tree level,
the Feynman diagrams contributing to $\pi\pi$ scattering amplitudes are presented in Fig.~\ref{pipitree}.
The corresponding invariant amplitude $A(s,t,u)$ reads:
\begin{equation}\label{Aamp}
	A(s,t,u)= -\frac{\lambda^2v^2}{9(s-m_\sigma^2)}- \frac{\lambda}{3}~.
\end{equation}
The invariant amplitudes $B(s,t,u)$ and $C(s,t,u)$ are related to $A(s,t,u)$ via crossing symmetry:
\begin{equation}\label{cs}
	A(s,t,u) = B(t,s,u)=C(u,t,s)~.
\end{equation}
From perturbative
unitarity in LSM we have:
\begin{equation}\label{Punit}
	\operatorname{Im}T^{0}_{0l}(s) ={\rho(s,m_\pi,m_\pi)}
	|T^{0}_{0t}(s)|^2~, \,\,\, {4m_\pi^2 < s < 4m_\sigma^2,}
\end{equation}
where  $T^{0}_{0t}(s)$ and $T^{0}_{0l}(s)$  denote the tree-level and
the one-loop  PWAs, respectively. Combining Eq~.\eqref{T0}-\eqref{pwj}, the  tree-level PWA is obtained:
\begin{equation}
	T^0_{0t}(s)=\frac{\lambda}{48\pi }\left(  \frac{ \left(3 m_\pi^2+2  m_\sigma^2-5 s\right)}{2 \left( s-m_\sigma^2\right)}+\frac{(m_\sigma^2- m_\pi^2) \log \left(\frac{s-4 m_\pi^2+ m_\sigma^2}{ m_\sigma^2}\right)}{ \left(s-4 m_\pi^2\right)}\right)~.
\end{equation}
At one-loop order,  it is tedious to present all Feynman diagrams and their corresponding results, which exceed 50 diagrams~\cite{Basdevant:1970nu}. Therefore, we will not include those amplitudes in this paper\footnote{The numerical code is available upon request.}.  With Eq.~\eqref{Punit},
it is easy to prove that the $[1,1]$ Pad\'e approximant
\begin{equation}\label{PadeA}
	T^{0[1,1]}_0(s) = \frac{T^{0}_{0t}(s)}{1 -
	T^{0}_{0l}(s)/T^{0}_{0t}(s)}~,
\end{equation}
satisfies elastic unitarity.

The $\sigma$ resonance corresponds
to the  pole of the PWA on the second Riemann sheet (RSII) of  complex $s$ plane, or the  zero of partial wave $S$
matrix:
\begin{equation}\label{Spipi}
	S(s) = 1 +2i\rho(s,m_\pi,m_\pi)T^{0[1,1]}_0(s)~,
\end{equation}
on the first Riemann sheet (RSI).

\begin{figure}[H]
	\begin{subfigure} {0.5\textwidth}
		\centering
		\includegraphics[width=1.\textwidth]{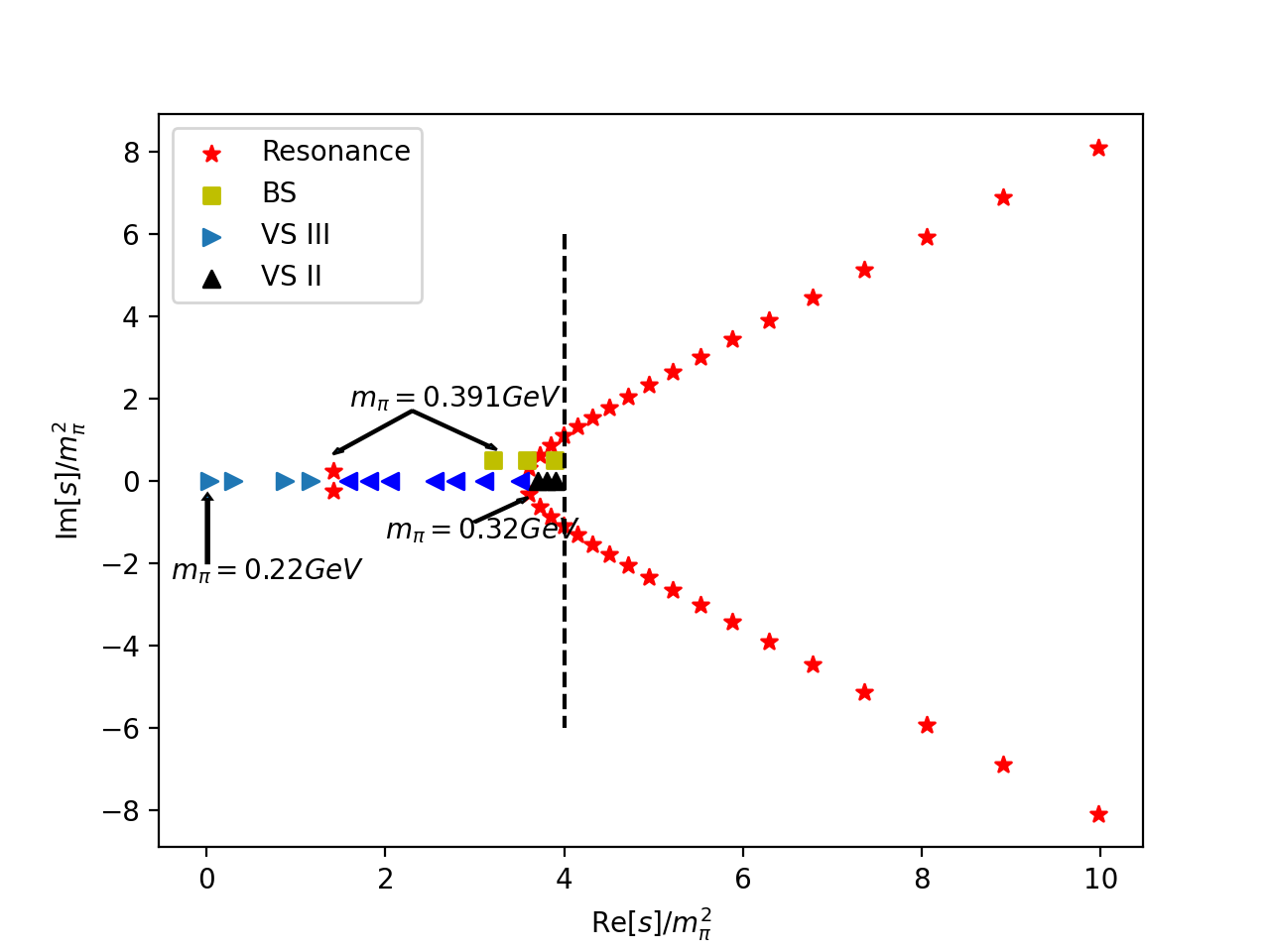}
	\end{subfigure}
	\begin{subfigure} {0.5\textwidth}
		\centering
		\begin{tikzpicture} [baseline=-3cm]
			\begin{feynhand}
				\vertex [NWblob](a){};
				\vertex [right =of a] (c);
				\vertex [above right =of c] (e);
				\vertex [below right =of c] (f);
				\vertex [left  =of a] (d);
				\vertex [above left =of d] (g);
				\vertex [below left =of d] (h);
				\propag[scalar] (c) to [edge label=$\pi$]            (e);
				\propag[scalar] (c) to  [edge label=$\pi$]           (f);
				\propag[scalar ] (a) to  [edge label=$\sigma$]           (d);
				\propag[scalar ] (c) to  [edge label=$\sigma$]           (a);
				\propag[scalar ] (h) to  [edge label=$\pi$]           (d);
				\propag[scalar ] (g) to      [edge label=$\pi$]       (d);
			\end{feynhand}
		\end{tikzpicture}
	\end{subfigure}
	\caption{The trajectory of $\sigma$ resonance with  $m_\pi$
		variation. The vertical dashed line denotes the physical threshold.  Right: the contribution of $\sigma$ self-energy correction to
		$\pi \pi$ amplitude.}
	\label{fig:fig-sig_v-png}
\end{figure}

{According to Eq.~\eqref{PadeA}, the numerator of $T^{0[1,1]}_{0}(s)$, i.e., $T^0_{0t}$,
contains a first-order pole at  $m_\sigma^2$.  When $m_\sigma>2m_\pi$, in the
denominator $1-T^{0}_{0l}/T^0_{0t}$, there also exists a
first-order pole because the  loop-level
amplitude $T^0_{0l}(s)$ contains a
second-order pole at $m_\sigma^2$ from the one-loop $\sigma$
propagator as shown in the right diagram of
Fig.~\ref{fig:fig-sig_v-png}. This causes
$T^{0[1,1]}_{0}(s)$ to be finite at $m_\sigma^2$. Thus, in this
situation $m_\sigma$ is not the pole mass of $\sigma$ and  the
$\sigma$ pole position would lie on the second Riemann sheet. On the contrary, with
$m_\pi$ growing up to  $2m_\pi>m_\sigma$, the second-order pole in
$T^0_{0l}(s)$ transforms to a first-order pole because the residue being
proportional to  $\Sigma(m_\sigma^2)$ equals  zero due to the
renormalization condition~\eqref{eq:Ren-cond-1}.  In this case, the
denominator of Eq.~\eqref{PadeA} is finite at $m_\sigma^2$, and the
numerator remains a pole at  $m_\sigma^2$ which corresponds to the $\sigma$
bound state.}

To obtain the specific trajectories through numerical calculation,
we choose a suitable  $m_\sigma=0.7\mathrm{GeV}$  such that
$\sigma$ pole  locates  at about $(0.47-i0.16)$~GeV at physical pion mass.\footnote{This width is slightly lower than  that given by PDG. In order to get a larger width, one can tune $m_\sigma=0.8-0.9\mathrm{GeV}$. But this choice is not satisfactory  in another aspect: The $\sigma$ only turns into a bound state when $m_\pi=0.4-0.45\mathrm{GeV}$, which  is somewhat larger than the result of Ref.~\cite{Cao:2023ntr}, where the bound state appears when $m_\pi < 0.391\mathrm{GeV}$ (see  Fig.~\ref{fig:fig-sig_v-png}).   Since in the present paper our purpose is only to obtain a qualitative picture of the pole trajectory, slightly different choices would not modify the main result of our paper.}  Then,
fixing this $m_\sigma$ parameter and
taking  into account the relation between $f_\pi$ and $m_\pi$,  the trajectory of $\sigma$ pole with increasing $m_\pi$ is depicted in
Fig.~\ref{fig:fig-sig_v-png}. The $\sigma$ resonance falls down  to
real axis below the threshold from the
complex plane above the threshold, becoming two virtual states (VSI and
VSII)  when $m_\pi$ increases from physical value to $m_\pi\simeq
	0.32\mathrm{GeV}$. One of them (VSII) runs
towards threshold and finally crosses the threshold to the real axis below
the threshold on RSI, turning into a bound state when $2m_\pi >
	m_\sigma$. On the other hand,
the other virtual state (VSI) runs away from the threshold and
collides with the third virtual state (VSIII) which appears from the
left-hand cut when $m_\pi\simeq 0.22\mathrm{GeV}$. Then these two
virtual-state poles turn into a pair of resonance poles
on the complex plane.
The trajectory is  similar to that of the Roy equation~\cite{Cao:2023ntr} and the  $N/D$ modified  $O(N)$ model~\cite{Lyu:2024lzr}, but now the critical point ($m_\sigma/2$)
when $\sigma$ becomes a bound state can be determined analytically
from the expression of Pad\'e amplitude. Note that appearance of VSIII is a result of the $\sigma$ exchange in the crossed channels which is the requirement of crossing symmetry. However, this effect is not taken into account in the unitarized $\chi$PT calculations and thus there is no such a virtual state found there~\cite{Hanhart:2008mx}.

Having examined that the $\sigma$ pole trajectory can be satisfactorily reproduced in the scheme
of linear $\sigma$ model with Pad\'e unitarization,
we are confident to step forward by studying
the $N^*(920)$ pole trajectory in the linear $\sigma$ model with nucleon field, in the next subsection.

\subsection{$S_{11}$ channel of  $\pi N$ scattering amplitude}
For the process $\pi^a(p)+N_i(q)\to \pi^{a^\prime}(p^\prime)+N_f(q^\prime)$,
the isospin amplitude can be decomposed as:
\begin{equation}
	T=\chi_f^{\dagger}\left(\delta^{a a^{\prime}} T^{+}+\frac{1}{2}\left[\tau^{a^{\prime}},
		\tau^a\right] T^{-}\right) \chi_i~,
\end{equation}
where  $\tau^a$ ($a=1,2,3$) are Pauli matrices, and $\chi_i$ ($\chi_f$) corresponds to the
isospin wave function of the initial (final) nucleon state. The
amplitudes with isospins $I = 1/2, 3/2$ can be written as
\begin{equation}
	\begin{aligned}
		 & T^{I=1 / 2}=T^{+}+2 T^{-}~, \\
		 & T^{I=3 / 2}=T^{+}-T^{-}~.
	\end{aligned}
\end{equation}
Taking into account of the Lorentz structure, for an isospin indices $I=1/2,3/2$, the amplitude can be represented as
\begin{equation}
	T^I=\bar{u}^{(s^{\prime})}\left(q^{\prime}\right)\left[A^I(s,
		t)+\frac{1}{2}\left(\slashed{p} +\slashed{p}^{\prime}\right)
		B^I(s, t)\right] u^{(s)}(q),
\end{equation}
with the superscripts $(s), (s^\prime)$ denoting the spins of Dirac
spinors and  three Mandelstam variables $s=(p+q)^2, t=(p-p^\prime),
	u=(p-q^\prime)$ obeying the constraint $s+t+u=2m_N^2+2m_\pi^2$. The
channel  with orbit angular momentum $L$, total angular momentum  $J$ and
total isospin  $I$ denoted as  $T(L_{2I2J})$ is defined as:
\begin{equation}
	T^{I,J}_{\pm}= T(L_{2I2J})=T^{I,J}_{++}(s) \pm T^{I,J}_{+-}(s),\quad L=J\mp
	\frac{1}{2},
\end{equation}
where the definition of partial wave helicity amplitudes are written as:
\begin{equation}
	\begin{aligned}
		{T}^{I,J}_{++}= & 2 m_N A^{I,J}_C(s
		)+\left(s-m_\pi^2-m_N^2\right) B^{I,J}_C(s)                             \\
		{T}^{I,J}_{+-}= & -\frac{1}{\sqrt{s}}\left[\left(s-m_\pi^2+m_N^2\right)
			A^{I,J}_S(s)+m_N\left(s+m_\pi^2-m_N^2\right) B^{I,J}_S(s)\right]
	\end{aligned}
\end{equation}
with
\begin{equation}\label{FSC}
	F_{C/S}^{I,J}(s)=\frac{1}{32\pi}\int_{-1}^1 \mathrm{~d} z_s F^I(s, t)\left[P_{J+1 /
				2}\left(z_s\right)\pm P_{J-1 / 2}\left(z_s\right)\right],\quad F=A,B
\end{equation}
$z_s=\cos\theta$ with  $\theta$ the scattering angle in center of mass  frame (CM) .  The PWAs $T^{I,J}_{\pm}$ satisfy unitarity condition:
\begin{equation}
	\operatorname{Im}T^{I,J}_{\pm}(s)
	=\rho(s,m_\pi,m_N)|T^{I,J}_{\pm}(s)|^2,\quad s>s_R=(m_\pi+m_N)^2\ .
\end{equation}

\begin{figure}[H]
	\centering
	\includegraphics[width=0.7\linewidth]{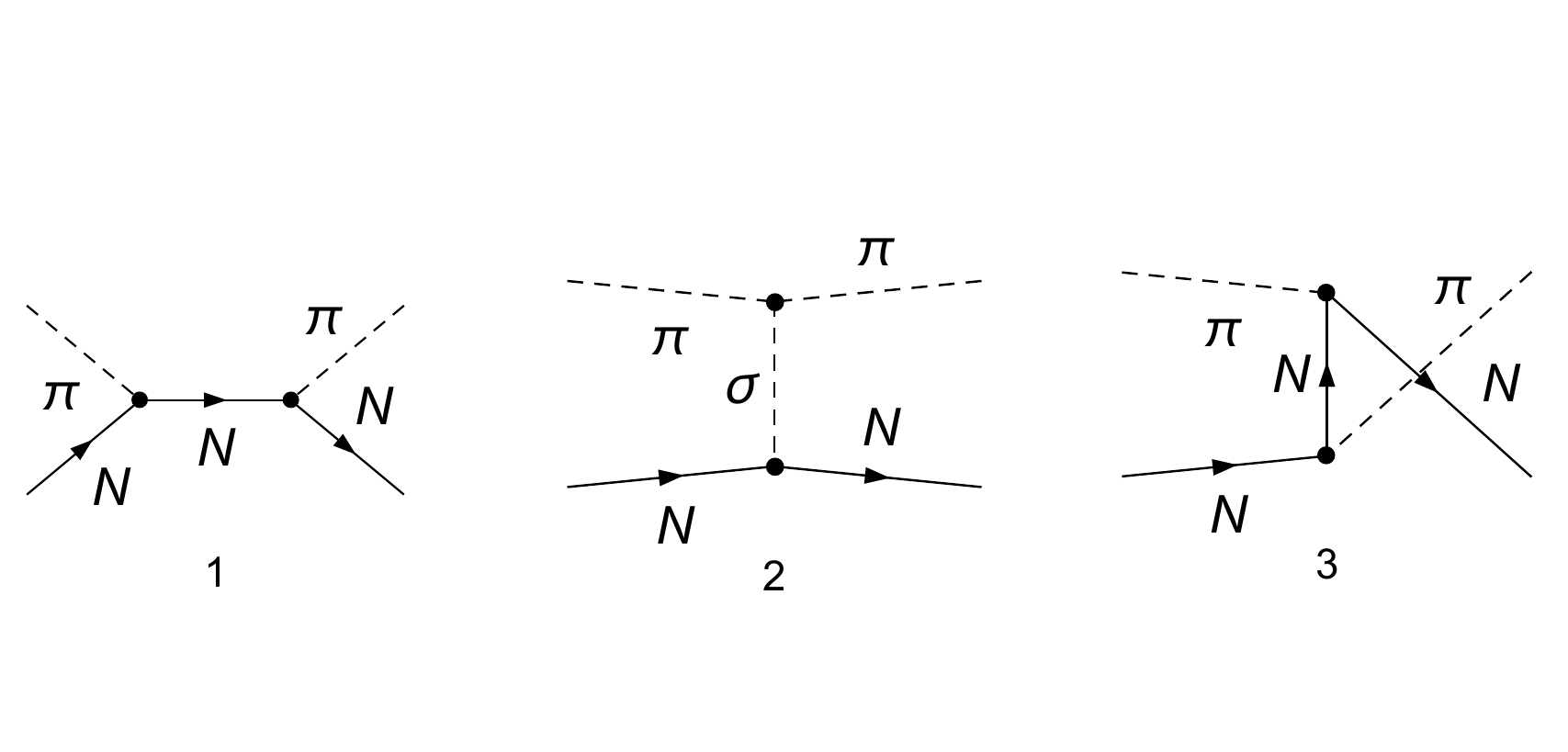}
	\caption{The tree-level Feynman diagrams contributing to $\pi N$ scatterings. }
	\label{piNT}
\end{figure}

The full tree-level amplitudes of $\pi N$ scatterings in LSM are given by three diagrams  as depicted in Fig.~\ref{piNT}.
Contributions to invariant amplitudes $A^{1/2}(s,t,u)$ and $B^{1/2}(s,t,u)$ at tree level read
\begin{equation}
	\begin{split}
		 & A^{1/2}(s,t,u)= -\frac{g\lambda v}{3(t-m_\sigma^2)}~,                  \\
		 & B^{1/2}(s,t,u)= -g^2\left(\frac{3}{s-m_N^2}+\frac{1}{u-m_N^2}\right)~.
	\end{split}
\end{equation}
According to Eq.~\eqref{FSC}, after partial-wave projection, the
expressions of $A^{1/2,1/2}_{C/S}$ and $B^{1/2,1/2}_{C/S}$  are listed
as follows,
\begin{equation}
	A_C^{1/2,1/2}(s)=-\frac{g\lambda v}{96\pi k^2}\left(1 - \frac{(m_\sigma^2+4k^2)I(s)}{4k^2}\right)~,
\end{equation}
\begin{equation}
	A_S^{1/2,1/2}(s)=-\frac{g\lambda v}{96\pi k^2}\left(1 - \frac{m_\sigma^2I(s)}{4k^2}\right)~,
\end{equation}
\begin{equation}
	B_C^{1/2,1/2}(s)=\frac{g^2}{32\pi}\left(-\frac{6}{s-m_N^2}+ \frac{1}{k^2} - \frac{m_N^2(s-c_L)\ln(\frac{s(s-c_R)}{m_N^2(s-c_L)})}{4sk^4}\right)~,
\end{equation}
\begin{equation}
	B_S^{1/2,1/2}(s)=\frac{g^2}{32\pi}\left(\frac{6}{s-m_N^2}+ \frac{1}{k^2} - \frac{(s-c_R)\ln(\frac{s(s-c_R)}{m_N^2(s-c_L)})}{4k^4}\right)~,
\end{equation}
with $k=\sqrt{s}\rho(s,m_\pi,m_N)/2$ being the magnitude of  3-momentum in CM and
\begin{equation}
	I(s)=\ln\left(\frac{\left((m^2_\pi-m_N^2)^2-2s(m_\pi^2+m_N^2)+s(s+m_\sigma^2)\right)}{m_\sigma^2s} \right)~.
\end{equation}
$B_C(s)$ and $B_S(s)$ contain  the $u$-cut in the interval $(c_L=(m_N^2-m_\pi^2)^2/m_N^2,c_R=m_N^2+2m_\pi^2)$   from the logarithmic term generated by $u$-channel nucleon exchange, as depicted   in the 3rd diagram  in Fig.~\ref{piNT}.
When $2m_\pi < m_\sigma < 2m_N$, the $I(s)$ function contains
circular arc cuts~\cite{Ma:2020sym} centered at the origin with a
radius of $m_N^2-m_\pi^2$. At one loop,  the circular cut emerges due to continuous two-particle spectrum,  which covers the circular arc cut.

After partial-wave projection, the perturbative PWAs will
be  unitarized by $N/D$ method~\footnote{
			Here $N/D$ method is used to avoid spurious poles which may present in the amplitude using Pad\'e approximation.
		}, which boils down to solving an
integral equation about  $N(s)$ function:
\begin{equation}
	N(s) = N(s_0) + U(s)-U(s_0) + \frac{(s-s_0)}{\pi} \int_{s_R}^{\infty}
	\frac{(U(s) -U(s^\prime))\rho(s^\prime,m_\pi,m_N)
		N(s^\prime)}{(s^\prime -s_0)(s^\prime -s)}ds^\prime~.
\end{equation}
The subtraction point $s_0$ and subtraction value  $N(s_0)$ can be
chosen appropriately and  $U(s)$ function should be analytic when
$s>s_R$, such that $N(s)$ only contains left hand
cuts
\begin{equation}
	U(s)-U(s^\prime)= \frac{s-s^\prime}{2\pi i} \int_L \frac{\operatorname{disc} M(\tilde{s})}{(\tilde{s}-s)(\tilde{s}-s^\prime)}\mathrm{d}\tilde{s}~,
\end{equation}
where the subscript $L$ denotes the left-hand cut where the integration is performed.  The
discontinuity of the
amplitude $M(s)$ need to be an input from  the
perturbative calculation. Since the dispersion relation of the
amplitude on the left-hand cut essentially gives the amplitude with the right-hand
cut integral subtracted up to a  polynomial, we can use the
perturbative amplitude with the right-hand cut dispersion integral
subtracted to estimate $U(s)-U(s')$ directly in the following.

The amplitude satisfying the unitarity condition can be constructed  as (we use $M(s)$ to represent $\pi N$ scattering amplitude in $S_{11}$ channel):
\begin{equation}
	\begin{split}
		 & M(s) = \frac{N(s)}{D(s)}~,                    \\
		 & D(s) = 1 - \frac{s-s_0}{\pi}\int_{s_R}^\infty
		\frac{\rho(s^\prime)N(s^\prime)}{(s^\prime-s)(s^\prime-s_0)}ds^\prime.
	\end{split}
\end{equation}
One can numerically solve the equation by inverse matrix
method, after introducing a cutoff $\Lambda$ such that the integral
interval becomes  $(s_R,\Lambda)$ instead of $(s_R,\infty)$. In the
following, $s_0$ and $\Lambda$,    are fixed at
$s_R$,  and $(m_N+m_\sigma)^2$, respectively.

\begin{figure}[H]
			\begin{subfigure}{1\textwidth}
				\centering
				\includegraphics[width=0.99\linewidth]{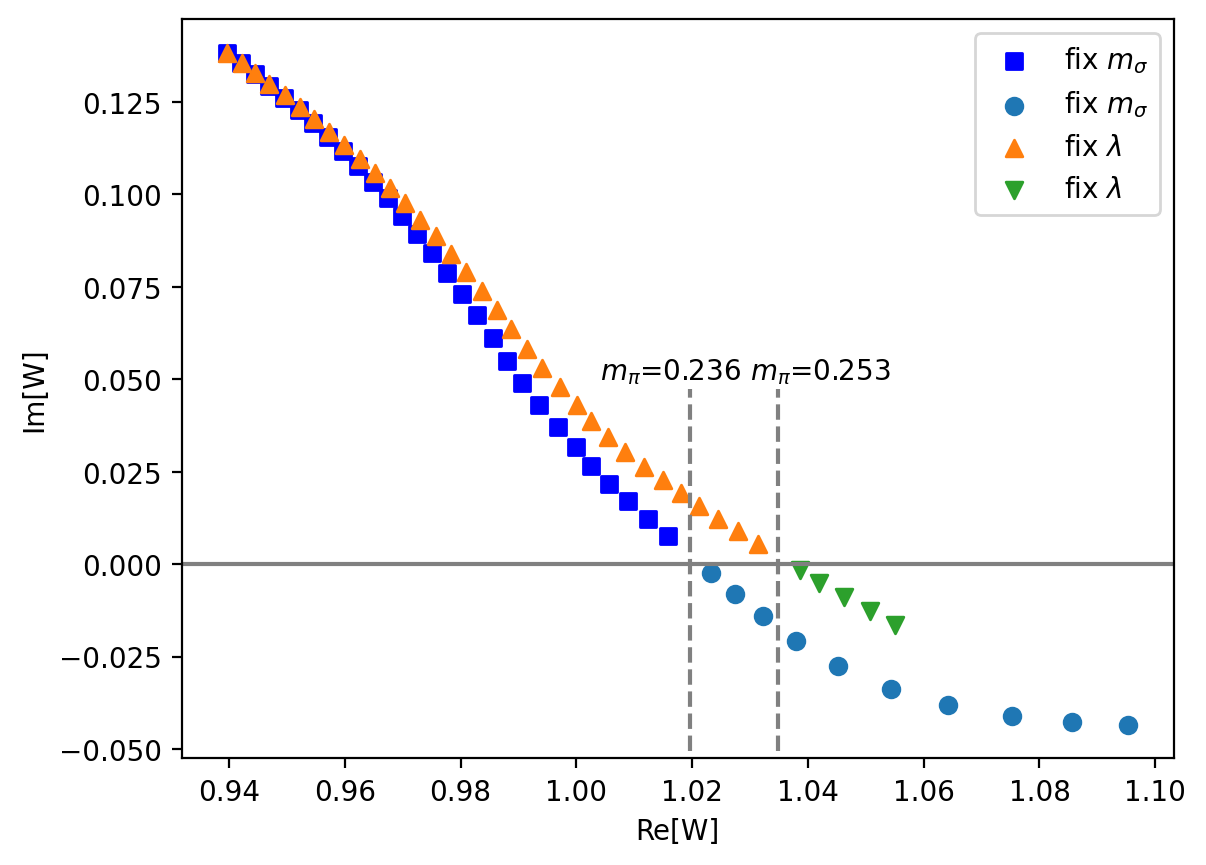}
			\end{subfigure}
			\caption{The trajectories of $N^*(920)$ as $m_\pi$ ranges from the physical value to $0.27\mathrm{GeV}$ for both cases of fixed $m_\sigma$ and fixed $\lambda$ at tree level. The points with $\operatorname{Im}[W]>0$ are the zeros (RSI)  of $S$ matrix for different pion masses and those with  $\operatorname{Im}[W]<0$  are the zeros of the $S_+$ function (see the text). The vertical dashed lines  correspond to  roots obtained by solving   equation~\eqref{eq:Imt-zero} for $m_\pi = 0.236\mathrm{GeV}$  and $0.253\mathrm{GeV}$, respectively. 
            }
			\label{fig:tree_zero}
		\end{figure}

	At tree level, since there is already no right-hand cut in tree-level amplitude $M_t$, we set $U(s)$ equal to
		$M_t(s)$, and
		$N(s_0)$ equal to $M_t(s_0)$. Starting from  $m_\sigma$ at $0.55\mathrm{GeV}$ \footnote{Here the $m_\sigma$ is chosen slightly different from $\pi\pi$ case, since the $N^*(920)$ pole is close to the Roy-equation result with this choice. }, we have performed  two sets of  calculations: one with fixed $m_\sigma$ and the other with fixed $\lambda$ as $m_\pi$ increases.  A pole at
		$(0.94 - 0.14i)\mathrm{GeV}$ corresponding  to $N^*(920)$ can be found on the second sheet with physical pion and nucleon mass.
		The pole trajectories on RSII with $m_\pi$ increasing for both cases are shown in  Fig.~\ref{fig:tree_zero} with $\operatorname{Im}W>0$ ($W\equiv\sqrt{s}$) points.
		The imaginary part of $N^*$ pole position decreases while
		the real part  grows when $m_\pi$ increases in both cases, causing the poles
		to move toward the  $u$-cut bounded by  two branch points at $c_R$
		and  $c_L$. Since the two conjugate $N^*$ poles on RSII correspond to zeros of the $S$ matrix on RSI, we consider the zero points in the following. After the zeros reach the $u$-cut, in principle there could be two possibilities: (1) the zeros move away from each other on the real axis after collision; (2) the zeros cross the $u$-cut and enter adjacent sheets defined by the logarithmic branch of the  $u$-cut.  
        However, we will show that the first senario is impossible for $\pi N$ scattering.

		In the first senario, if the  $N^*$ zeros reach the $u$-cut at $s_* \in(c_L,c_R)$ at critical pion mass $\tilde{m}_\pi$ and then separate to two virtual state zeros, 
		this implies that $s_*$ must be a second-order zero of the $S$ matrix. A necessary condition is that $s_*$ is also a second-order zero of the imaginary part of $S$ matrix, i.e., $\operatorname{Im}[S(s_*)] = 0$ and $\operatorname{Im}[S'(s_*)] = 0$. The first condition gives
        \begin{equation}2i\rho(s_*,m_\pi,m_N) \operatorname{Im}[M(s_*)]=0~,
       \end{equation}     
        as $2i\rho(s_*,m_\pi,m_N)$ is real on the $u$-cut.
        Since the sole contribution to the imaginary part of PWA  arises from the $u$-channel nucleon pole $\frac {g}{u-m^2_N}$ in the partial wave projection, which is non-perturbative and  remains valid  to all orders of perturbative expansions~\footnote{The circular cut does  not affect the  imaginary part of PWA on the $u$-cut despite intersecting it.}, the zero of the imaginary part of PWA on the $u$-cut in general case always coincides with tree level $\mathrm {Im}M_t(s)$.  
 		According to the expression of $M_t(s)$, the condition  $\operatorname{Im} M_t(s) = 0$ on the $u$-cut is equivalent to
		\begin{equation}
			(m_N - m_\pi - W) (m_N + m_\pi - W) [m_N (m_N - W) (m_N + W)^2-m_\pi^4]=0~, \quad  W\equiv \sqrt s.
			\label{eq:Imt-zero}
		\end{equation}
		This equation has   only one solution at $W_*\simeq m_N-\frac{m_\pi^4}{4m_N^3}$ inside  $u$-cut.
          
        Similarly,  $\operatorname{Im}[S'(s_*)] = 0$ requires that the derivative of the left-hand side of    Eq.~\eqref{eq:Imt-zero} vanishes at the same point. However, this is impossible: differentiating the   left-hand  side of   Eq.~\eqref{eq:Imt-zero} yields $m_N (m_N - 3 W) (m_N + W)$ (only consider the term in brackets), whose only reasonable root  $W^\prime_*=m_N/3$ does not satisfy  Eq.~\eqref{eq:Imt-zero}. Therefore, the nonexistence  of  a second-order zero  and only one zero in imaginary part of PWA excludes  the first scenario.
 
         In the context of the $N/D$ method, the zero point of the $S$ matrix at $s_*$ requires
            \begin{equation}
			1+ 2i\rho(s_*,m_\pi,m_N) M(s_*)= \frac{D(s_*)+2i\rho(s_*,m_\pi,m_N)N(s_*)}{D(s_*)}=0~.
		\end{equation}
		Since $D(s)$ function is real  on the left-hand cuts, the requirement    $\operatorname{Im} S = 0$  leads to
		\begin{equation}
			2i\rho(s_*,m_\pi,m_N) \operatorname{Im}[N(s_*)]=2i\rho(s_*,m_\pi,m_N) \operatorname{Im}[M_t(s_*)]=0~,
		\end{equation}
		where we have used the fact that $2i\rho(s_*,m_\pi,m_N)$ is real.
		Thus, the necessary condition for the $S$ matrix to vanish at $s_*$ is $\operatorname{Im} M_t(s_*) = 0$ which is reduced to \eqref{eq:Imt-zero}.
	Note that the property that the  zero of the imaginary part of $S$ matrix on the $u$-cut coincides with  that of Im$M_t(s)$  is  not only preserved in $N/D$ method but also respected by Pad\'e or  $K$-matrix unitarizations. A proof is provided in    App.~\ref{zeroImT}.
	
\begin{figure}[H]
			\centering
			\includegraphics[width=0.8\linewidth]{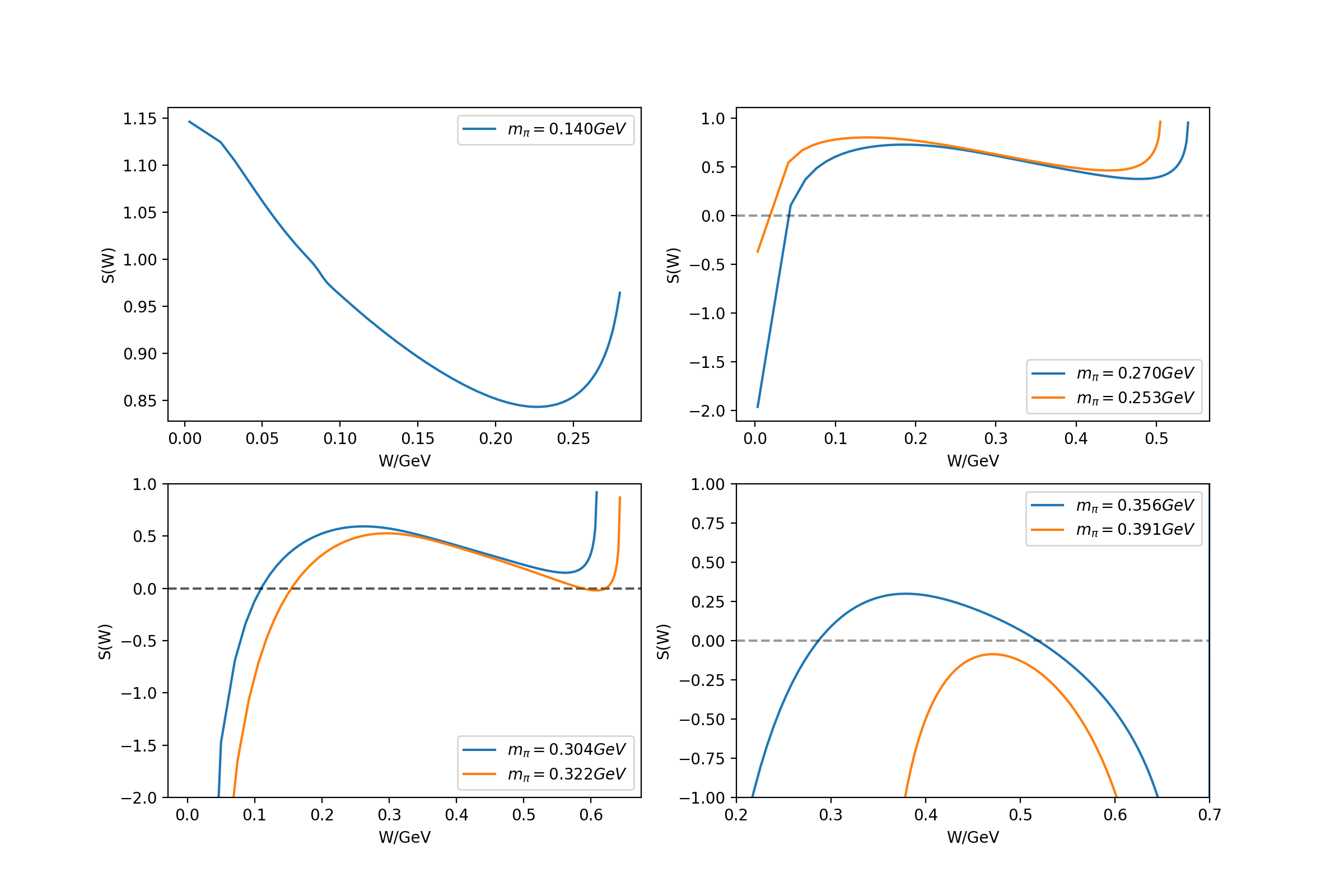}
			\caption{The $S$ matrix values for $\pi\pi$ scatterings between the left hand cut and  the threshold using the results of the previous section. 
           Initially, no $S$-matrix zeros exist in this region (upper-left subfigure). As $m_\pi$ increases, a virtual-state zero (VSIII) emerges from the left-hand cut (upper-right). The sigma resonance turns into virtual-state zeros when the $S$ matrix's local minimum contacts the real axis, generating a second-order zero precisely at this point. Subsequently, the two virtual-state zeros (VSI and VSII) split apart along the real axis (lower-left). VSI later becomes a bound-state pole of the $S$-matrix, leaving two virtual-state zeros (VSII and VSIII). These two zeros then coalesce on the real axis and move into the complex plane as resonance zeros (lower-right). 
            }
			\label{fig:pipism}
		\end{figure}

        It is, however, worth mentioning that the first scenario we have just excluded is actually what was happening in the case  of $\pi\pi$  scatterings (see Fig.~\ref{fig:pipism}).  

		Having excluded  Case 1, the only possibility is that the $N^*$ zeros cross the $u$-cut and  enter adjacent Riemann sheets defined by the $u$-cut. To trace the zeros further, the  $S$ matrix must be analytically continued onto these sheets. The analytically continued  $S$ matrix from the upper (lower) half-plane  to the adjacent lower (upper) half-plane, $S_+$ ($S_-$), can be obtained by adding  $\pm 2\pi i$ to  the logarithmic  function $\ln(\frac{s(s-c_R)}{m_N^2(s-c_L)})$ in the unitarized $S$ matrix on RSI.
Therefore, after $N^*$  crosses the $u$-cut, we should find the zero of $S_+$ on the adjacent lower half-plane or the zero of $S_-$ on the upper adjacent half-plane to continuously trace the trajectories. Fig~.\ref{fig:tree_zero} shows how the zeros approach the $u$-cut, cross it at the critical  point ( marked by vertical dashed lines  in Fig.~\ref{fig:tree_zero}) where $\operatorname{Im}[S]$ vanishes, and  subsequently become the zeros of $S_+$ as $m_\pi$ increase for both schemes: fixed $m_\sigma$ and fixed $\lambda$. 

        Notably, tracing the poles across the $u$-cut starting from   RSII is consistent with tracing the  zeros  starting from  RSI. The $S^{II}$ ($S$ matrix on the RSII) is defined 
        \begin{equation}
            S^{II}(s)= \frac{1}{S(s)}~.
        \end{equation}
        Modifying  the logarithmic  function $\ln(\frac{s(s-c_R)}{m_N^2(s-c_L)})$ in $S^{II}(s)$ similarly results in 
        \begin{equation}
            S^{II}_{\pm}(s) =  \frac{1}{S_\pm(s)}~.
        \end{equation}
        Therefore, the zeros of $S_\pm$ correspond to  the poles of  $S^{II}_{\pm}$.

		  In above, we have actually established a mathematical theorem: \emph{if the zero reaches the $u$-cut, it must meet the cut at $W=W_*$ defined by Eq.~\eqref{eq:Imt-zero}, and will cross onto the adjacent sheet. }  Nevertheless, this theorem does not tell why the pole should move toward the real axis. Then the next question becomes whether the zero will ever approach the $u$-cut and why?   At  tree level, the $N^*$ resonance does reach the $u$-cut. However, there may still exist other possibilities for the $N^*$ pole of the full amplitude: it could move onto the real axis within the interval $(c_R,s_R)$ outside the $u$-cut, or it may even never touch the real axis. To  further explore these possibilities, we extend the calculations to one-loop level in the following. 

At one-loop level, parameter $N(s_0)$ is set  equal
to  $M_t(s_0)+M_l(s_0)$ and $U(s)$ is  written as:
\begin{equation}
	U(s) = M_t(s) + M_l(s) - \frac{s}{\pi}\int_{s_R}
	\frac{\rho(s^\prime,m_\pi,m_N)M^2_t(s^\prime)}{s^\prime(s^\prime-s)}~,
\end{equation}
with $M_l(s)$ the one-loop correction to the  PWA. The full one-loop amplitude has been known for a long time~\cite{Mignaco:1971ys}, and after partial wave projection, the PWA is too long to be presented here~\footnote{The code is also available upon request.}. The third term in the above expression ensures  $U(s)$ to be analytic in the interval
$(s_R,(m_N+m_\sigma)^2)$ and  consistent with perturbative unitarity:
\begin{equation}
	\operatorname{Im}M_l(s) = \rho(s,m_\pi,m_N) |M_t(s)|^2,\quad  s >
	s_R~.
\end{equation}
The nearest inelastic cut in the one-loop amplitude is above $(m_\sigma+m_N)^2$ in real axis, and thus we
fix  the cutoff  $\Lambda=(m_N+m_\sigma)^2$ such that the unitarized
amplitude satisfies the single channel unitary condition in the  interval
$(s_R,(m_N+m_\sigma)^2)$.

\begin{figure}[H]
	\begin{subfigure}{0.5\textwidth}
		\centering
		\includegraphics[width=0.8\linewidth]{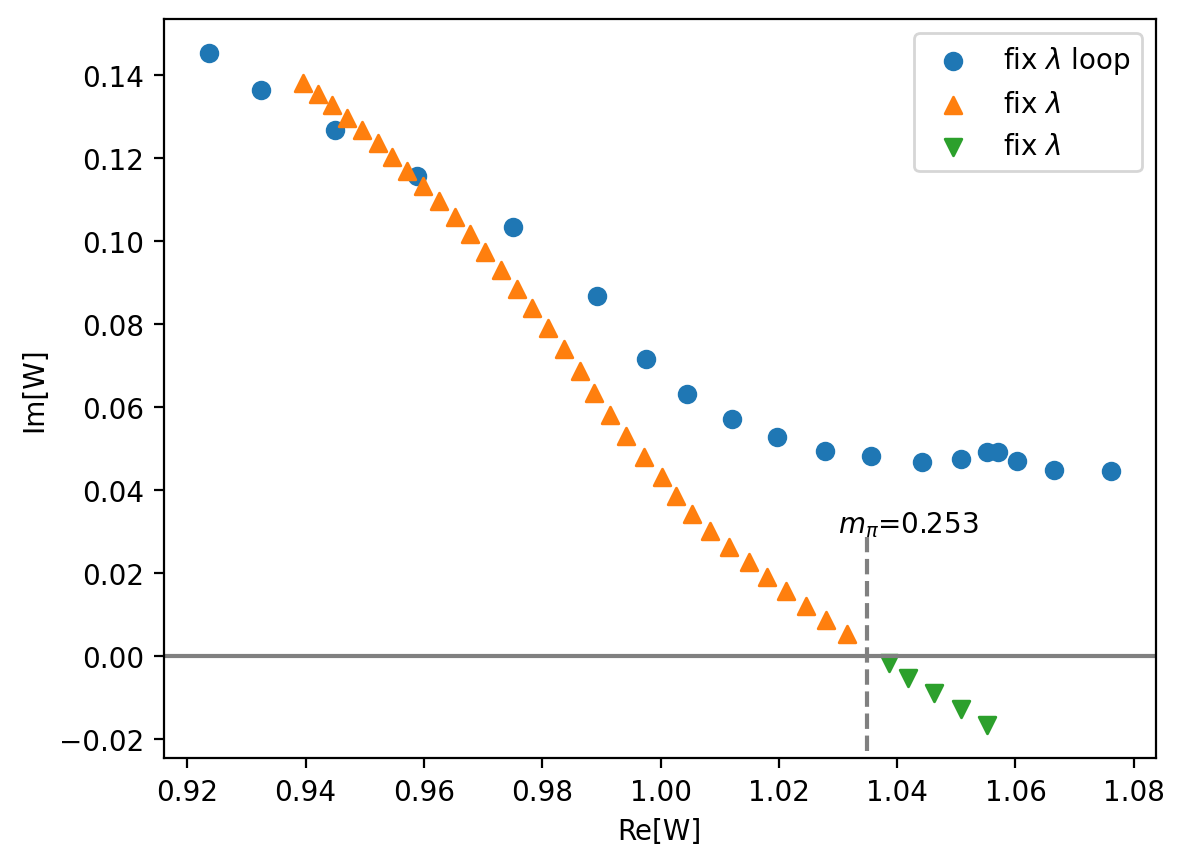}
	\end{subfigure}
	\begin{subfigure}{0.5\textwidth}
		\centering
		\includegraphics[width=0.99\linewidth]{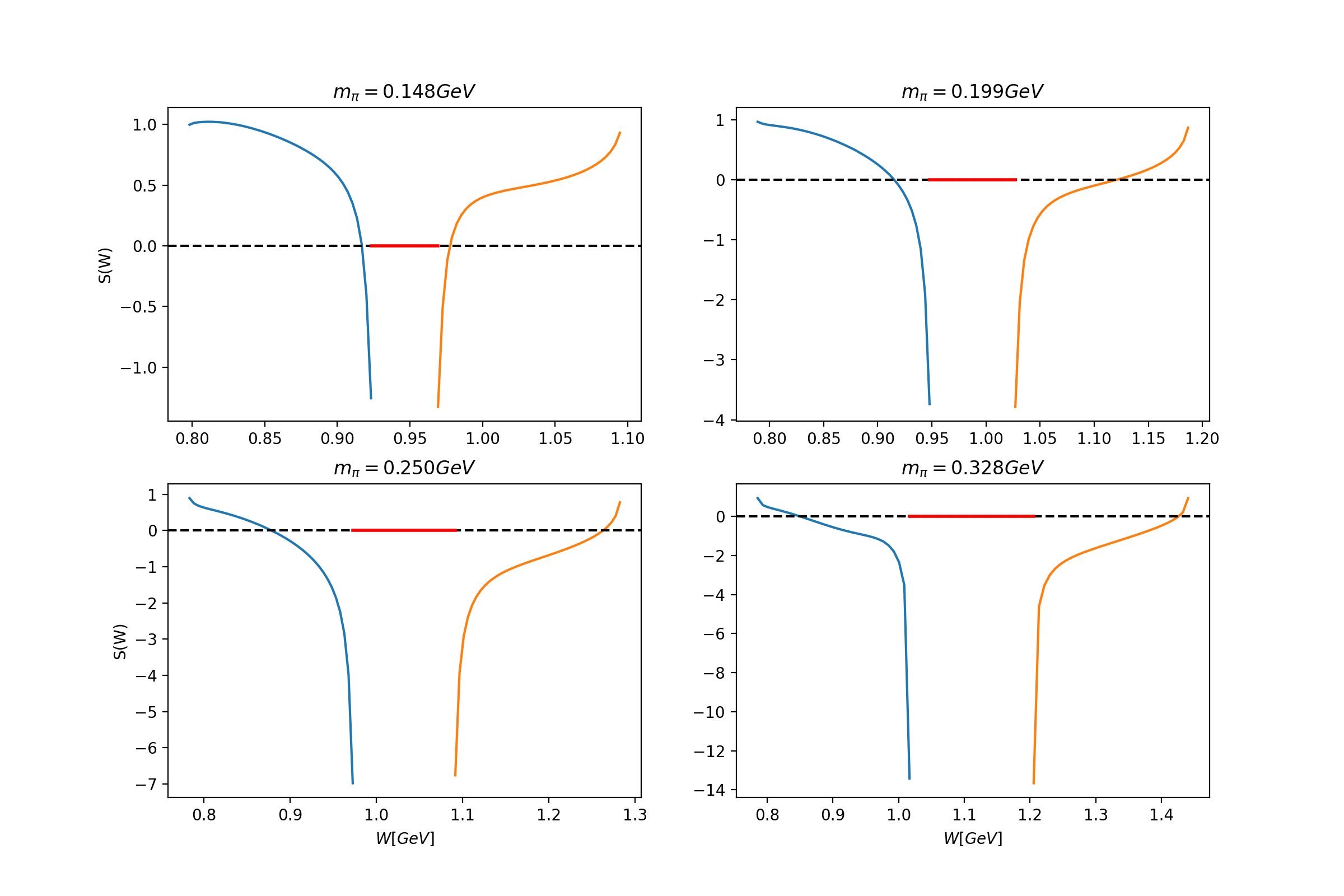}
	\end{subfigure}
	\caption{ Left:
		The dot points  and triangle points denote the trajectories of $N^*(920)$  with  $m_\pi$ variation from $0.138\mathrm{GeV}$ to $0.360\mathrm{GeV}$, at  one-loop level and tree level  for fixed $\lambda$, respectively. 
		  Right: the values of  $S$
		matrix at one-loop level in intervals   $(s_L,
			c_L)$ and    $(c_R,s_R)$ on the first Riemann sheet. The red lines denote  $u$-cuts and
		endpoints of black lines are  $s_L$ and  $s_R$, respectively.
		The blue intersection points represent virtual states. 
		\label{fig:fig-tree_zero_var-png}
	}
\end{figure}
A pole {at  $(0.92 - 0.15i)\mathrm{GeV}  $}  corresponding to $N^*(920)$ can be found  on the second sheet
	with physical pion and nucleon masses.
The pole trajectories of $N^*(920)$ for the tree level and the one-loop
level  as $m_\pi$ increases are shown in
Fig.~\ref{fig:fig-tree_zero_var-png}. 
At one-loop level, the trend  of $N^*(920)$ trajectories is similar to that at tree level: both move toward the real axis in the beginning.   However, when $m_\pi$ gets larger, the difference emerges. The one-loop zero does not seem likely to fall onto the real axis at least up to $m_\pi=0.36\mathrm{GeV}$. We do not further test larger $m_\pi$ since chiral expansions, which constrains the $m_\pi$ dependence of $m_N$ and $f_\pi$ in our calculations, may become more and more inaccurate at larger $m_\pi$.

We also studied the behavior of the two virtual states~\cite{Li:2021oou}  lying in the
interval $( s_L=(m_N-m_\pi)^2,c_L)$ and  $(c_R,s_R)$, respectively.
The values of  $S$ matrix calculated with one-loop level input in the two intervals on
the first Riemann sheet  are
plotted in Fig.\ref{fig:fig-tree_zero_var-png}, which demonstrates the
fact that the $S$ matrix equals to unity at  $s_L$ and  $s_R$ by
definition while it tends to negative infinity when $s$ is close to the
two branch points  $c_L$ and  $c_R$. As a  result, a zero point inside each
interval occurs.  The calculation reveals that the virtual states move toward
$s_R$ or  $s_L$ with increasing $m_\pi$.
\section{Discussions and Conclusions}\label{disc}

In this paper we have studied the $\sigma$ pole trajectory and the $N^*(920)$ pole trajectory
with varying $m_\pi$, in a linear $\sigma$ model with nucleons, aided
by  certain unitarization approximations.  The $m_\pi$ dependence
of $f_\pi$ and $m_N$ from the $\chi$PT are also taken into account, which
renders the LSM more reasonable in approximating the low energy QCD.
The $\sigma$ pole trajectory is found to be in agreement with previous
studies~\cite{Cao:2023ntr,Lyu:2024lzr}. The result on  $N^*(920)$
pole
trajectory is novel. At tree level, the $N^*(920)$ pole is found to move towards the
$u$-cut on the real axis on the second Riemann sheet with increasing
$m_\pi$,  ultimately crossing the $u$-cut and entering the adjacent Riemann sheet defined by the $u$-cut. At one-loop level, however,   the $N^*(920)$ still stays on the complex-plane  at $m_\pi=0.36\mathrm{GeV}$ and even higher values. The intricate    analytical structure  of $\pi N$ PWAs, including the circular cut and $u$-cut,  which stems from the the dynamical complexity of $\pi N$ interactions,  increase the difficulty  for predictions on  the final fate of $N^*$.
The results presented in this paper can be useful in comparison with future lattice studies on $\pi N$ scatterings.
 The next interesting topic for future studies would be to investigate the $N^*(920)$ pole trajectory in the presence of temperature and chemical potential, and
the old concept of parity doublet model may return with some new ingredients.

\begin{acknowledgments}
	This work is supported by National Natural Science Foundation of China
	under Contract No. 12335002, 12375078.
	This work is also supported by “the Fundamental Research Funds for the Central Universities”.
\end{acknowledgments}
\appendix
\newpage
\section{The expressions of counter terms and renormalization constants}\label{ece}
The counter terms  and renormalization constants take following forms up to one-loop level,
\begin{flalign}
	\begin{split}
		 & Z_\phi  =1 -\frac{ g^2}{4 \pi ^2} \left(B_0\left(m^2_\pi,m_N^2,m_N^2\right)+m_\pi^2B^\prime_0\left(m_\pi^2,m_N^2,m_N^2\right)\right)-\frac{\lambda^2v^2}{144\pi^2}B^\prime_0\left(m_\pi^2,m_\pi^2,m_\sigma^2\right),
	\end{split} &
\end{flalign}
\begin{flalign}
	\begin{split}
		Z_F = & 1-\frac{g^2}{32 \pi ^2 m_N^2}\left(3 m_\pi^2 B_0\left(m_N^2,m^2_\pi,m_N^2\right)+m_\sigma^2 B_0\left(m_N^2,m_N^2,m_\sigma^2\right)-3 A_0\left(m^2_\pi\right)+4 A_0\left(m_N^2\right)-A_0\left(m_\sigma^2\right)\right) \\
		      & +\frac{g^2}{16 \pi ^2 }\left(3m^2_\pi B^\prime_0\left(m_N^2,m_\pi^2,m_N^2\right)+m_\sigma^2B^\prime_0\left(m_N^2,m_N^2,m_\sigma^2\right)-4m_N^2B^\prime_0\left(m_N^2,m_N^2,m_\sigma^2\right)\right)~,
	\end{split} &
\end{flalign}
\begin{flalign}
	\begin{split}
		Z_g= & 1 + \frac{g^2}{16 \pi ^2 m_N^2}\left(-3  m^2_\pi B_0\left(m_N^2,m^2_\pi,m_N^2\right)-m_\sigma^2 B_0\left(m_N^2,m_N^2,m_\sigma^2\right)+3  A_0\left(m^2_\pi\right)-4 A_0\left(m_N^2\right)\right.                     \\
		     &\left.+A_0\left(m_\sigma^2\right)\right) + \frac{g^2}{16 \pi ^2}\left(2  B_0\left(m_N^2,m_N^2,m_\sigma^2\right)+3m^2_\pi B^\prime_0\left(m_N^2,m^2_\pi,m_N^2\right)+m_\sigma^2B^\prime_0\left(m_N^2,m_N^2,m_\sigma^2\right)\right.\\
           & \left. -4m_N^2B^\prime_0\left(m_N^2,m_N^2,m_\sigma^2\right)\right)~,
	\end{split} &
\end{flalign}
\begin{flalign}
	\begin{split}
		\delta m_\pi^2 = & \frac{\lambda^2 v^2}{144 \pi ^2}\left( B_0\left(m_\pi^2,m_\pi^2,m_\sigma^2\right) - m_\pi^2 B^\prime_0\left(m_\pi^2,m_\pi^2,m_\sigma^2\right)\right)+\frac{\lambda}{96 \pi ^2}\left( 5  A_0\left(m^2_\pi\right)+ A_0\left(m_\sigma^2\right)\right) \\
		                 & -\frac{g^2}{4\pi^2} \left(m_\pi^4B^\prime_0\left(m_\pi^2, m_N^2, m_N^2\right)+2A_0(m_N^2)\right)~,
	\end{split} &
\end{flalign}
\begin{flalign}
	\begin{split}
		\delta m_\sigma^2= & \frac{\lambda^2v^2 }{\pi^2} \left(\frac{\operatorname{Re}  B_0\left(m_\sigma^2,m^2_\pi,m^2_\pi\right)}{96}+ \frac{B_0\left(m_\sigma^2,m_\sigma^2,m_\sigma^2\right)}{32}-\frac{m_\sigma^2B^\prime_0\left(m^2_\pi,m^2_\pi,m_\sigma^2\right)}{144}\right) \\
      &  +\frac{\lambda}{32\pi^2}\left(A_0\left(m^2_\pi\right)+A_0\left(m_\sigma^2\right)\right)
		                    -\frac{g^2m_\sigma^2}{4 \pi ^2}\left(B_0\left(m^2_\pi,m_N^2,m_N^2\right)-  B_0\left(m_\sigma^2,m_N^2,m_N^2\right)\right.\\
                            &\left. +m^2_\pi  B^\prime_0\left(m_\pi^2,m_N^2,m_N^2\right)\right)                                               -\frac{g^2}{4\pi^2}\left(2 A_0\left(m_N^2\right) +4 m_N^2  B_0\left(m_\sigma^2,m_N^2,m_N^2\right)
		\right)~.
	\end{split} &
\end{flalign}
The definitions of 1-point function $A_0(m^2)$  and 2-point function $B_0(p^2,m_1^2,m_2^2)$ are expressed as
\begin{equation}
	\begin{split}
		 & A_0(m^2)\equiv -16\pi^2i\int \frac{\mathrm{d}^4k
		}{(2\pi)^4} \frac{1}{k^2-m^2}~,                                  \\
		 & B_0(p^2,m^2_1,m_2^2)\equiv -16 \pi^2i\int \frac{\mathrm{d}^4k
		}{(2\pi)^4} \frac{1}{(k^2-m_1^2)[(p+k)^2-m_2^2]}~.
	\end{split}
\end{equation}
$B_0^\prime$ denotes the derivation with respect to the first argument.
\section{The zero of imaginary part of  $S$ matrix in other unitary models}\label{zeroImT}

In [1,1] Pad\'e  approximant, the ampltiude $M(s)$
\begin{equation}
	M(s) = \frac{M^2_t(s)}{M_t(s)-M_l(s)}~.
\end{equation}
 The fact that   $\mathrm{Im}[M_l(s)]=\mathrm{Im} [M_t(s)]$ on the   $u$-cut  causes the zero of Im$M(s)$ to be the same as that of Im$M_t(s)$ on the cut. As to the  $K$-matrix, the amplitude  is given  by\footnote{Using $i\rho(s,m_\pi,m_N) $ instead of $\tilde{B}(s)$ leads to same result.}
\begin{equation}
	M^{-1}(s)=M^{-1}_t(s) - \tilde{ B}(s),\quad \tilde{B}(s) = b_0 + \frac{s-s_*}{\pi} \int_{s_R} \frac{\rho(s^\prime,m_\pi,m_N)}{(s^\prime-s_*)(s^\prime-s)}ds^\prime~.
\end{equation}
Then, the  Im$S(s)$ on the $u$-cut reads:
\begin{equation}
	\begin{split}
		\operatorname{Im}S(s) & = \operatorname{Im}\left[ \frac{1 -\tilde{B}(s)M_t(s)+2i\rho(s,m_\pi,m_N)M_t(s)}{1-\tilde{B}(s)M_t(s)}\right]                      \\
		                      & =\frac{2i\rho(s,m_\pi,m_N)\operatorname{Im}M_t(s)}{1-\tilde{B}(s)(M_t^*(s)+M_t(s))+\tilde{B}^2(s)|M_t(s)|^2},\quad s\in(c_L,c_R)~.
	\end{split}
\end{equation}
Again, we get the same conclusion.
\bibliographystyle{apsrev4-2}
\bibliography{ref}

\end{document}